\documentclass[12pt]{iopart}
\usepackage{xcolor}
\usepackage{graphicx}

\newcommand{\Neel}{N\'{e}el } 

\newcommand{\bn}{\mathbf{n}}
\newcommand{\bM}{\mathbf{m}}
\newcommand{\dnt}{ \dot{\mathbf{n}} }
\newcommand{\dMt}{ \dot{\mathbf{m}} }
\newcommand{\ddnt}{ \ddot{\mathbf{n}} }

\newcommand{\bmA}{\mathbf{M}_1}
\newcommand{\bmB}{\mathbf{M}_2}

\newcommand{\btM}{\bM}

\newcommand{\mbr}{\mathbf{r}}
\newcommand{\bd}{\mathbf{\dmid}}
\newcommand{\bj}{\mathbf{j}}

\newcommand{\lt}{\exch}

\newcommand{\funWM}{\frac{ \delta f}{ \delta \bM} }
\newcommand{\funWn}{\frac{ \delta f}{ \delta \bn} }

\newcommand{\ez}{  \hat{z} }
\newcommand{\ex}{  \hat{x} }

\newcommand{\parnxii}{\frac{\partial \bn}{\partial \xi_i}}

\newcommand{\dxit}{\dot{\xi}_i}

\newcommand{\ddxit}{\ddot{\xi}_i}

\newcommand{\bR}{\mathbf{R}}

\newcommand{\intrhot}{\int_0^\infty d\rhot}

\newcommand{\rhot}{\tilde{\rho}}
\newcommand{\parThRhot}{\frac{ \partial \theta }{ \partial \rhot }}
\newcommand{\parThRhoRhot}{\frac{ \partial^2 \theta }{ \partial \rhot^2 }}

\newcommand{\bBA}{\mathbf{B}_1}

\newcommand{\exch}{H_{\mathrm{exc}}}
\newcommand{\iExch}{A}
\newcommand{\anis}{H_{\mathrm{an}}}
\newcommand{\dmid}{d}
\newcommand{\dmiD}{D}
\newcommand{\h}{H}
\newcommand{\bh}{\mathbf{H}}
\newcommand{\xo}{x_0}
\newcommand{\Ms}{M_s}
\newcommand{\Gn}{\alpha_G}

\newcommand{\Ca}{C_1}
\newcommand{\Cb}{C_2}
\newcommand{\Cc}{C_3}

\begin{document}

\title[Antiferromagnetic skyrmion motion]{Phenomenology of current-induced skyrmion motion in antiferromagnets}
\author{H. Velkov$^{1,2}$, O. Gomonay$^{1,3}$, G. Schwiete$^{1}$ and J. Sinova$^{1,4}$}

\address{ 
$^1$ Institut f\"{u}r Physik, Johannes Gutenberg-Universit\"{a}t Mainz, 55128 Mainz, Germany
}
\address{
$^2$ Graduate School Materials Science in Mainz, Staudinger Weg 9, 55128 Mainz, Germany
}
\address{
$^3$ National Technical University of Ukraine “KPI”, 03056, Kyiv, Ukraine
}
\address{
$^4$ Institute of Physics ASCR, v.v.i., Cukrovarnicka 10, 162 53 Praha 6 Czech Republic
}
\author{A. Brataas$^5$}

\address{ 
$^5$ Department of Physics, Norwegian University of Science and Technology, NO-7491 Trondheim, Norway
  }
\vspace{10pt}
\author{M. Beens$^6$ and R. A. Duine$^{6,7}$}

\address{ 
$^6$ Institute for Theoretical Physics and Center for Extreme Matter and Emergent Phenomena, Utrecht University, Leuvenlaan 4, 3584 CE Utrecht, The Netherlands
  }
\address{
$^7$ Department of Applied Physics, Eindhoven University of Technology, PO Box 513, 5600 MB, Eindhoven, The Netherlands
}
\ead{
hvelkov@uni-mainz.de
}

\vspace{10 pt}

\begin{abstract}

We study current-driven skyrmion motion in uniaxial thin film antiferromagnets in the presence of the Dzyaloshinskii-Moriya interactions and in an external magnetic field. We phenomenologically include relaxation and current-induced torques due to both spin-orbit coupling and spatially inhomogeneous magnetic textures in the equation for the \Neel vector of the antiferromagnet. Using the collective coordinate approach we apply the theory to a two-dimensional antiferromagnetic skyrmion and estimate the skyrmion velocity under an applied DC electric current.
\end{abstract}


%
%
%

%
%
%
\maketitle
%
%

\section{Introduction}

For decades ferromagnets have been the main components of spintronic devices \cite{Zutic2004}. Antiferromagnets (magnetically ordered materials with compensated magnetization) have long remained in their shadows \cite{Sinova2012, Jungwirth2016}, even though they possess properties that make them appealing as potential alternatives of ferromagnets and as next generation data storage devices. The insensitivity of antiferromagnets to external magnetic fields makes them more robust against magnetic perturbations, they operate on faster timescales possibly enabling ultrafast information processing, and antiferromagnetic metals, alloys and semiconductors are not limited to combinations of only a few elements like Fe, Co, Ni \cite{Gomonay2014}. On the other hand, the insensitivity to external magnetic fields makes antiferromagnets also much harder to manipulate and control \cite{Jungwirth2016}. In recent years, several breakthroughs were achieved in overcoming this obstacle. The anisotropic magnetoresistance (AMR) effect was proposed \cite{Shick2010} and utilized \cite{Marti2014} to electrically detect antiferromagnetically ordered states. Another important step towards antiferromagnetic spintronics was the prediction \cite{Zelezny2014} and subsequent observation \cite{Wadley2016} of \Neel spin-orbit torques in a certain class of antiferromagnets that can electrically manipulate the antiferromagnetic \Neel vector. 


Independently of the developments with antiferromagnets, skyrmions have been gaining momentum in the field of data storage \cite{Nagaosa2013}. Magnetic skyrmions, for example, are topological windings of the magnetization on the nanoscale that appear in noncentrosymmetric materials. In ferromagnets, skyrmions are promising candidates as information carriers for future information-processing devices and have been intensely studied in recent years \cite{Nagaosa2013, Fert2013, Sampaio2013, Iwasaki2013a, Iwasaki2013b}. One of the key properties that makes them attractive is the very small current densities that are needed to set them in motion \cite{Fert2013}.

In contrast, not much is known about skyrmions in antiferromagnets. While their existence has been predicted by Bogdanov and colleagues \cite{Bogdanov2002} (see also \cite{Ivanov1995}), there are no experimental observations. The differences and possible advantages of antiferromagnets over ferromagnets lead to the question: how will skyrmions interact with currents in antiferromagnets?

To utilize skyrmions, it is necessary to know how to create, manipulate and detect them in magnetic thin-film nanostructures. Initial numerical analyses have been performed on two-dimensional antiferromagnetic films \cite{Zhang2015b, Barker2015} focusing on the manipulation of skyrmions by electric currents. In both treatments the skyrmion dynamics were studied by solving the equations of motion for the two sublattices numerically where damping and current-induced torques were implemented in the same spirit as for ferromagnets. In this article, we use a phenomenological and analytical approach to gain further insights.


The latter analytical approach has been applied in recent works to study the magnetization dynamics of antiferromagnets under the influence of electric currents \cite{Gomonay2010, Hals2011, Tveten2013, Cheng2014c}. However, so far the effect of the inversion symmetry breaking inducing the Dzyaloshinskii-Moriya (DM) interactions has not been taken into account. Here, starting from general principles and symmetry considerations, we construct the equations describing the macroscopic antiferromagnetic magnetization dynamics in the presence of a DC current and an external magnetic field. In particular, we i) take into account the effect of the DM interactions corresponding to a particular symmetry class of magnetic materials, ii) incorporate phenomenologically the spin-orbit torques of an antiferromagnetic system and iii) study the role of the external magnetic field and its effect on the skyrmion motion. We demonstrate that the latter has, in fact, no effect on the skyrmion motion. It does, however, affect its shape.

\begin{figure}[t!]
	\centering
		\includegraphics[scale=0.4]{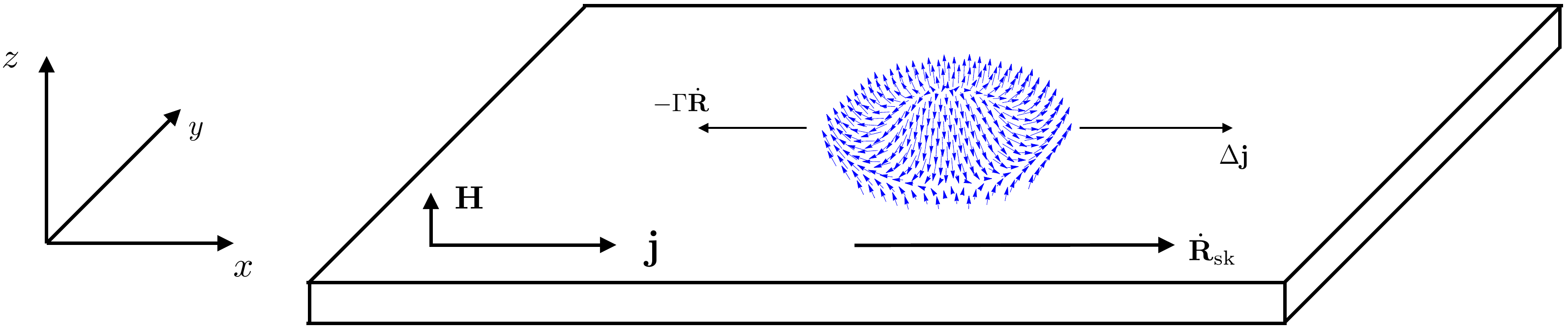}
	\caption{Schematic representation of the skyrmion motion in an antiferromagnet driven by an electric current $\bj \parallel \ex$ and an external magnetic field $\bh \parallel \ez$, as described by (\ref{eq:final_result_skyrmion_motion}). The friction force is denoted by $\Gamma \dot{\bR}$ and the longitudinal current-induced force by $\Delta \bj$. The combination of these forces leads to a skyrmion motion with the velocity $\dot{\bR}_{\mathrm{sk}}$. For simplicity, a static skyrmion shape is depicted, without taking into account the shape changes that the magnetic field and the electric current induce. }
	\label{fig:skyrmion_motion}
\end{figure}

Our main result is an equation of motion for the position $\bf{R}$ of the skyrmion in the thin film given by
\begin{eqnarray}
	m_{\mathrm{eff}}
	\ddot{\bR}
	&=
	-
	\Gamma 
	\dot{\bR}
	+
	\Delta
	\bj.
\label{eq:final_result_skyrmion_motion}
\end{eqnarray}
Here, $m_\mathrm{eff}$ is the effective skyrmion mass, and $\Gamma$ a friction constant. The external magnetic field does not appear in the equation. The dissipative spin-orbit and spin-transfer torques (described by $\Delta$) lead to a longitudinal current-induced force on the skyrmion. In figure \ref{fig:skyrmion_motion} we illustrate the resulting skyrmion motion. We emphasize that we included both homogeneous and inhomogeneous current-induced torques in our analysis, thus obtaining a more general form of the equation of motion than considered in \cite{Zhang2015b, Barker2015}. 

The remainder of the paper is structured as follows. In section \ref{section:phenomenology}, we discuss the phenomenological model and the resulting equations describing the current-induced antiferromagnetic dynamics. We discuss the current-induced torques and we derive a closed equation of motion for the antiferromagnetic order parameter. In section \ref{section:skyrmion_motion} we reformulate the result into an equation for the position of a skyrmion by using the collective coordinate approach. Finally, we provide an estimate for the skyrmion velocity.

\section{Phenomenological model}
\label{section:phenomenology}

In this section we analyze the general magnetization dynamics of the antiferromagnets of interest. We first discuss the magnetic energy of the model and then derive the equation of motion for the \Neel vector.

\subsection{Energy functional}

We consider a two-sublattice antiferromagnet within the exchange approximation with the sublattice magnetizations $\bmA \simeq -\bmB$ and $\left| \bmA \right| = \left| \bmB \right| = \Ms$, where $\Ms$ is the saturation magnetization. Further, we consider a noncentrosymmetric lattice of the crystallographic class $C_{n v}$ \cite{Bogdanov2002}. This class encompasses also two-dimensional films which have structural inversion asymmetry along the $\ez$-direction, for example, due to the presence of an interface.

It is most convenient to formulate the theory in terms of the antiferromagnetic order parameter (also called the N\'{e}el vector) $\bn = (\bmA - \bmB)/2 \Ms$ and the total magnetization $\bM = (\bmA + \bmB)/2 \Ms$. Our phenomenological approach is based on the exchange approximation \cite{Landau1997}, which requires rotational invariance of the magnetization vectors and invariance of the theory with respect to an exchange of the two sublattices, i.e., under the transformations $\bn \rightarrow - \bn$ and $\bM \rightarrow \bM$. The magnetic energy that follows from these considerations is\footnote{The energy functional seems to satisfy the $C_{\infty v}$ symmetry group, however, the discrete symmetry origin of the antiferromagnetic vectors needs to be respected, which lowers the symmetry to $C_{nv}$. } \cite{Bogdanov2002, Landau1997} 
\begin{eqnarray}
	F
	=
	\int d\mbr
	\Bigg\{
	&\iExch 
		\sum_{i={x,y}}
		\left( 		\partial_i \bn		\right)^2
		+
		\exch \Ms \bM^2
		-
		\anis \Ms
		( \bn \cdot \ez )^2
		-
		2 \Ms \bM \cdot \bh
\nonumber \\
	&+
		2
		\Ms
		\bd \cdot ( \bM \times \bn )
		+
		\dmiD
		\Big[
			( \ez \cdot \bn)
			( \nabla \cdot \bn)
			-
			( \bn \cdot \nabla )
			( \ez \cdot \bn )
		\Big]
	\Bigg\}.
\label{eq:afm_free_energy}
\end{eqnarray}
The first and the second terms describe the inhomogeneous and homogenous exchange interaction with the constants $\iExch$ and $\exch$, respectively. The uniaxial anisotropy is parameterized by the constant $\anis>0$ and the external magnetic field is denoted by $\bh$.  The remaining terms describe the DM interactions, where $\bd \parallel \ez$ and $\dmiD$ represent the homogeneous and inhomogeneous parts, respectively. The latter part of the DM interactions, also called a Lifshitz invariant, is the main ingredient needed to stabilize a skyrmion in this system and inhomogeneous textures in general \cite{Bogdanov2002}.

The antiferromagnetic vectors obey the constraints $\bn^2 = 1$ and $\bn \cdot \bM = 0$. Throughout this work we make the assumption that the homogeneous exchange interaction $\exch$ is the dominant energy scale, so that  $\exch \gg \dmid, \anis$. For typical values of the external magnetic field that do not destroy the antiferromagnetic order the exchange constants dominates the field too, $\exch \gg \h$ \cite{Marti2014}.

\subsection{Landau-Lifshitz-Gilbert equations}
\label{section:LLG_eqs}
In this section we derive the equations of motion for the \Neel order parameter for a time-independent magnetic field and electric current.
The Landau-Lifshitz-Gilbert equations to leading order in $\exch$ are given by \cite{Gomonay2010, Hals2011, Baryakhtar1979}
\begin{eqnarray}
	\dnt 		&\simeq 	- \frac{\gamma}{2 \Ms} \Bigg[ 	\bn \times \funWM 							\Bigg] 	+ \tau_{\bn}, 									\nonumber \\
	\dMt 		&\simeq 	- \frac{\gamma}{2 \Ms} \Bigg[ 	\bn \times \funWn   	+ \bM \times \funWM			\Bigg] 	+ \Gn \bn \times \dnt 		+ \tau_{\bM},
\label{eq:LLG_eqs}
\end{eqnarray}
where $\gamma$ is the gyromagnetic ratio, $\Gn$\footnote{In the notation of \cite{Hals2011}, this is the $G_2$ damping constant.} is a phenomenological Gilbert damping coefficient and $\tau_{\bn, \bM}$ represent the respective current-induced torques to be discussed in section \ref{section:current-induced_torques}.  The functional derivatives of the energy density $f$ are
\begin{eqnarray}
	\frac{1}{2 \Ms}
	\funWn
	&=
	- \frac{\iExch}{\Ms} \nabla^2 \bn
	+\bd \times \bM
	- \anis (\bn \cdot \ez) \ez
	+\frac{\dmiD}{\Ms}
	\left[
		(\nabla \cdot \bn) \ez
		- 
		\nabla (\bn \cdot \ez)
	\right]
\nonumber \\
	\frac{1}{2 \Ms}
	\funWM
	&=
	\exch \bM
	-
	\bh
	+
	( \bn \times \bd ).
\label{eq:afm_functional_derivatives}
\end{eqnarray}

The static magnetization of this system without electric currents is found by taking the cross product of the first line of (\ref{eq:LLG_eqs}) with $\bn$ and identifying the time-independent components under the constraints $\bn^2=1$ and $\bn \cdot \bM = 0$, which yields \cite{Bogdanov2002, Ivanov1995}
\begin{equation}
	\bM_0 
	=
	-
	\frac{1}{\exch}
	\bn \times (\bn \times \bh)
	-
	\frac{1}{\exch}
	\bn \times \bd.
\label{eq:static_equilibrium_magnetization}
\end{equation}
The presence of the homogeneous DM interactions term $d$ shows that even in the absence of an external magnetic field the total magnetization is non-zero. However, this term does not contribute to the motion of an antiferromagnetic texture within the approximations we consider and we neglect it for the remainder of the paper.

\subsection{Current-induced torques}
\label{section:current-induced_torques}

We follow a phenomenological approach to derive the current-induced torques. It is based on the Onsager reciprocity relations, which, in the case under consideration, relate the process of inducing charge currents by a time-varying magnetic texture to the effect that charge currents have on the magnetization dynamics. 

Following \cite{Hals2011}, to lowest order in the spatial gradients and the magnetization $\bM$ and zeroth order in spin-orbit coupling we find three contributions  towards the magnetically pumped charge density $\bj^\mathrm{pump} / \sigma$ that obey the symmetries of the system (rotation and $\bn \rightarrow - \bn$): $\eta \Ms/\gamma \, \bn \cdot ( \dMt \times \partial_i \bn)$, $\beta \Ms/\gamma \, \dnt \cdot \partial_i \bn$ and $\zeta \Ms/\gamma \, \bn \cdot (\dnt \times \partial_i \bM)$ \cite{Hals2011}. After applying the Onsager relations these are transformed and result in the torques
\begin{eqnarray}
	\tau_{\bn,\mathrm{STT}} 		&=		\frac{\eta}{2}   (\bj \cdot \nabla) \bn 		
							+
							\frac{\beta}{2}   (\bj \cdot \nabla) \bn \times \bM
							+
							\frac{\zeta}{2}  \bM \times \Big[ (\bj \cdot \nabla) \bM \times \bn 	\Big],
																							\nonumber \\
	\tau_{\bM,\mathrm{STT}}		&=		\frac{\eta}{2}  \bM \times \Big[ (\bj \cdot \nabla) \bn \times \bn 		\Big]
							+
							\frac{\beta}{2} (\bj \cdot \nabla) \bn \times \bn 
							+
							\frac{\zeta}{2} \bn \times \Big[ (\bj \cdot \nabla) \bM \times \bn 	\Big].
																							\nonumber \\	
\label{eq:STT_terms}
\end{eqnarray}
Here, the terms parameterized by the coefficients $\eta, \zeta$ describe reactive spin-transfer torques, whereas the term with $\beta$ describes a dissipative spin-transfer torque. 

Furthermore, the inversion symmetry breaking gives rise to another set of torques even in homogeneous systems, which do not involve gradients of the antiferromagnetic vectors. Following the same approach as above \cite{Duine2016}, we find the pumped charge currents that are lowest order in the spin-orbit coupling: $\Ca \Ms/\gamma \, \ez \times \dMt$, $\Cb \Ms/\gamma \, \ez \times \left( \bn \times \dnt \right)$ and $\Cc \Ms/\gamma \, \ez \times \left( \bM \times \dMt \right)$, which after applying the Onsager relations lead to 
\begin{eqnarray}
	\tau_{\bn,\mathrm{SOT}} 		&=			- \frac{\Ca}{2} \bn \times (\ez \times \bj)
								- \frac{\Cb}{2} \bM \times \Big[ 	\bn \times (\ez \times \bj) 	\Big]
								- \frac{\Cc}{2} \bn \times  \Big[	\bM \times (\ez \times \bj) 	\Big],
																							\nonumber \\
	\tau_{\bM,\mathrm{SOT}}		&=			- \frac{\Ca}{2} \bM \times (\ez \times \bj)
								- \frac{\Cb}{2} \bn \times  \Big[ \bn \times (\ez \times \bj) 	\Big]
								- \frac{\Cc}{2} \bM \times \Big[ \bM \times (\ez \times \bj) 	\Big].
																							\nonumber \\
\label{eq:SOT_terms}
\end{eqnarray}
Here, the terms proportional to $\Ca$ are field-like (reactive) spin-orbit torques and the terms proportional to $\Cb, \Cc$ are the anti-damping (dissipative) spin-orbit torques. In ferromagnetic systems similar torques have been discussed in \cite{Knoester2014, Hals2014} and in antiferromagnets a microscopic analysis has been performed in \cite{Zelezny2014}. 

We emphasize that the spin-orbit torques and the spin-transfer torques differ in their nature. Whereas in the latter the free electrons are polarized by the local magnetization while moving through the texture and interact with it after being polarized, in the spin-orbit torques the polarization is due to the spin-orbit coupling in the system and not due to the magnetization. In that sense, the spin-transfer torques are a result of a non-local interaction between the electrons and the magnetic moments, while the spin-orbit torques are local.

In the later steps of the calculation, presented below, the form of both the spin-transfer and the spin-orbit torques will be simplified by retaining only the leading order terms in $\exch$.


\subsection{Equation of motion}
Writing out the torques explicitly, the Landau-Lifshitz-Gilbert equations (\ref{eq:LLG_eqs}) become 
\begin{eqnarray}
	\dnt 		&= 	- \frac{\gamma}{2 \Ms} \Bigg[ 	\bn \times \funWM 						\Bigg] 						+ \frac{\eta}{2}   (\bj \cdot \nabla) \bn			- \frac{\Ca}{2} \bn \times (\ez \times \bj),			\nonumber \\
	\dMt 		&= 	- \frac{\gamma}{2 \Ms} \Bigg[ 	\bn \times \funWn 	+ \bM \times \funWM		\Bigg] 	+ \Gn \bn \times \dnt 		+ \frac{\eta}{2} \bM \times \Big[ (\bj \cdot \nabla) \bn \times \bn \Big]
\nonumber \\
			&\quad \, \,
																							+ \frac{\beta}{2} (\bj \cdot \nabla) \bn \times \bn	
																		- \frac{\Ca}{2} \bM \times (\ez \times \bj) 
																		- \frac{\Cb}{2} \bn \times  \Big[ \bn \times (\ez \times \bj) 	\Big].		
\label{eq:LLG_eqs_final}
\end{eqnarray}
Here, only terms to leading order in $\exch$ have been kept, apart from the $\eta$ and $\Ca$ ones in the second line, which we retained in order to keep the constraints $\bn^2 = 1$ and $\bn \cdot \bM = 0$ fulfilled. The term containing the functional derivative of the energy density with respect to $\bM$ is of subleading order in $\lt$, however, it is of the same order as the left hand side after substituting (\ref{eq:total_magnetization_expression}) below and needs to be kept as well. 

An expression for the total magnetization can be obtained from the equation for $\dnt$:
\begin{eqnarray}
	\bM
	=
	\frac{1}{\gamma \lt}
	\bn \times \dnt
	+
	\btM_0
	-
	\frac{\eta}{2 \gamma \lt}	
	\bn \times (\bj \cdot \nabla) \bn
	+
	\frac{\Ca}{2 \gamma \lt}
	\bn \times \left[ \bn \times (\ez \times \bj) \right],
\nonumber \\
\label{eq:total_magnetization_expression}
\end{eqnarray}
where $\btM_0$ now does not contain the homogeneous DM interactions contribution $\bd$, as discussed in section \ref{section:LLG_eqs}. With this, we are able to write a closed equation for the \Neel order parameter 
\begin{eqnarray}
	\bn \times \ddnt
	=
	\bh_{\mathrm{shape}}
	+
	\bh_{\mathrm{forces}}.
\label{eq:final_AFM_equation_of_motion}
\end{eqnarray}
Here, we have grouped the right hand side into terms that determine the shape of the antiferromagnetic texture and terms that induce or affect its motion. The fields are given by 
\begin{eqnarray}
	\bh_{\mathrm{shape}} 
	&=
	\frac{ 2 \gamma }{ \Ms } \dnt ( \bn \cdot \bh_{\mathrm{eff}})
	-
	\frac{\gamma^2 \lt}{2 \Ms^2}
	\left[
		\bn \times \funWn 
	\right]
	-
	\frac{\gamma^2}{\Ms^2}
	(\bn \times \bh_{\mathrm{eff}}) 
	(\bn \cdot \bh_{\mathrm{eff}}),
\nonumber \\
	\bh_{\mathrm{forces}}
	&=
	-
	\frac{\gamma \lt \eta}{2 \Ms}	
	\bn
	\Big[ 
		(\bj \cdot \nabla) \bn \cdot \btM
	\Big]
	+
	\frac{\gamma \lt \beta}{ 2 \Ms }
	(\bj \cdot \nabla) \bn \times \bn
\nonumber \\
	&\quad \,
	-
	\frac{\gamma \lt \Cb}{ 2 \Ms } 
	\bn \times
	\Big[
		\bn \times (\ez \times \bj)
	\Big]
	+
	\frac{ \gamma \lt \Gn }{ \Ms }
	\bn \times \dnt
\end{eqnarray}
and $\bh_{\mathrm{eff}} = \bh - C_1 \Ms / 2 \gamma (\ez \times \bj)$.

In deriving the equation for the \Neel order parameter only terms up to linear order in $\dnt$ and $\bj$ have been kept. An example of a term of higher order that has been omitted is $\bn \times ( \bj \cdot \nabla ) \dnt$. 

Equation (\ref{eq:final_AFM_equation_of_motion}) is an important result of this work and describes the magnetization dynamics of a uniaxial antiferromagnet with inversion symmetry broken along the $\ez$-direction under the influence of an external time-independent magnetic field and DC electric current.

\section{Skyrmion motion} 
\label{section:skyrmion_motion}

In the previous section we analyzed the magnetization dynamics of a uniaxial antiferromagnet in the presence of electric currents and a time-independent external magnetic field. Here, we focus on the translational motion of a magnetic skyrmion, rewrite the equations of motion by using the collective coordinate approach, and obtain an estimate for the skyrmion velocity as a result of the electric currents.

\subsection{Collective coordinates}
Previous work has shown that the dynamics of magnetic textures in ferromagnets and antiferromagnets can often be described by only a few variables \cite{Ivanov1995, Tveten2013}.  The approach necessitates the choice of a finite set of collective coordinates $\xi_i(t)$ which are used to specify the time evolution of the \Neel order parameter $\bn(\mbr, t) = \bn (\mbr, \{ \xi_i (t) \})$. In particular, we use 
\begin{eqnarray}
	\dnt &= \sum_i \parnxii \dxit,							\nonumber \\
	\ddnt &= \sum_i \parnxii \ddxit + \mathcal{O} (\dxit^2),
\label{eq:collective_coordinates_ansatz}
\end{eqnarray}
where the second term in the last equation is neglected because it is quadratic in the velocities, which are assumed to be small \cite{Tveten2013}. 

Here, we apply the approach to the translational motion of an antiferromagnetic skyrmion to analyze its current-induced dynamics. We assume that the skyrmion profile is composed of a static, cylindrical and rigid component $\bn_{\mathrm{sk}}$ and motion- and current-induced corrections $\delta \bn$ that break the cylindrical symmetry (see section \ref{sec:skyrmion_profile}). For the time evolution we use the ansatz \mbox{$\bn (\mbr, t) = \bn (\mbr - \bR(t))$}, where $\bn = \bn_{\mathrm{sk}} + \delta \bn$ and $\bR(t)$ is the skyrmion position. As collective coordinates we take $\{ \xi_i \} = \{R_x, R_y\}$. After multiplying (\ref{eq:final_AFM_equation_of_motion}) by $\bn \times \partial \bn / \partial x_\alpha$\footnote{The choice of this factor comes from general considerations for the conserved quantities in the system. For the translational motion, the relevant quantity is the momentum \cite{Ivanov1995}. Note that this also agrees with the approach taken in \cite{Tveten2013}.} for $\alpha=x,y$ and integrating over space, the equation of motion for the skyrmion position to leading order in the electric currents becomes
\begin{eqnarray}
	m_{\mathrm{eff}}
	\ddot{\bR}
	&=
	-
	\Gamma 
	\dot{\bR}
	+
	\Delta
	\bj.
\label{eq:final_skyrmion_collective_coordinate_eom}
\end{eqnarray}
The coefficients read 
\begin{eqnarray}
	m_{\mathrm{eff}}	 		&= 			\frac{ \Ms^2}{ \gamma^2 } \frac{ \xo }{ \exch },									\nonumber \\
	\Gamma					&\simeq 		- \frac{ \Ms}{ \gamma } \xo \, \Gn,										\nonumber \\
	\Delta					&\simeq 		\frac{ \Ms }{ 2 \gamma }
										\left[
											\beta \xo
											-
											\Cb \xo^2 I
										\right],
\label{eq:force_language_final_skyrmion_eom_coefficients}
\end{eqnarray}
where $\Gamma$ represents a friction term and $\Delta$ characterizes the effect of the dissipative current-induced torques. The dimensionless constant $ I $ is determined by the skyrmion profile, which we discuss later. The characteristic lengthscale $\xo$ is the domain wall width of the system and is given in the following section. The dependence of the effective mass $m_{\mathrm{eff}}$ on the exchange constant $\exch$ is the main difference compared to ferromagnetic skyrmion motion and results from the different nature of the magnetization dynamics in antiferromagnets. 

In deriving (\ref{eq:final_skyrmion_collective_coordinate_eom}) we have considered both homogeneous and inhomogeneous current-induced torques, as well as an external magnetic field applied in the $\ez$-direction. Our results thus paint a richer picture of the current-induced antiferromagnetic skyrmion motion than discussed recently. References \cite{Zhang2015b, Barker2015} predicted longitudinal current-induced forces on the skyrmion position, as opposed to the ferromagnetic skyrmion motion, where the nonzero magnetization always leads to a transverse force. Reference \cite{Zhang2015b} dealt with homogeneous torques only, whereas in \cite{Barker2015} only gradient torques have been considered. In both references no magnetic field has been included.  We find, however, that even in the presence of an applied field the skyrmion motion remains longitudinal. This is further substantiated by the findings of \cite{Ivanov1995, Ivanov1994}, where it is shown that gyroscopic forces (which are linear in the magnetic field) are not present in antiferromagnets for objects that exhibit the topology of skyrmions.

For an electric current applied in the $\ex$-direction, an expression for the longitudinal velocity $v_{\mathrm{sk}}$ can be readily obtained by the steady-state solution of (\ref{eq:final_skyrmion_collective_coordinate_eom}), which yields
\begin{eqnarray}
	v_{\mathrm{sk}} 
	= 
	-
	\Bigg[
		\frac{ \beta	 	}{ 2 \Gn }
		-
		\frac{ \Cb \xo	}{ 2 \Gn }
		I
	\Bigg]
	j_x
	\equiv
	v_{\mathrm{sk},\beta}
	+
	v_{\mathrm{sk},\Cb}
	.
\label{eq:longitudinal_skyrmion_velocity}
\end{eqnarray}
This is the velocity corresponding to the zero-field scenario considered in \cite{Zhang2015b, Barker2015}. Before we proceed with its estimate, we need to analyze the skyrmion shape in more detail.

\subsection{Skyrmion profile}
\label{sec:skyrmion_profile}

\begin{figure}[t!]
	\centering
		\includegraphics[scale=0.6]{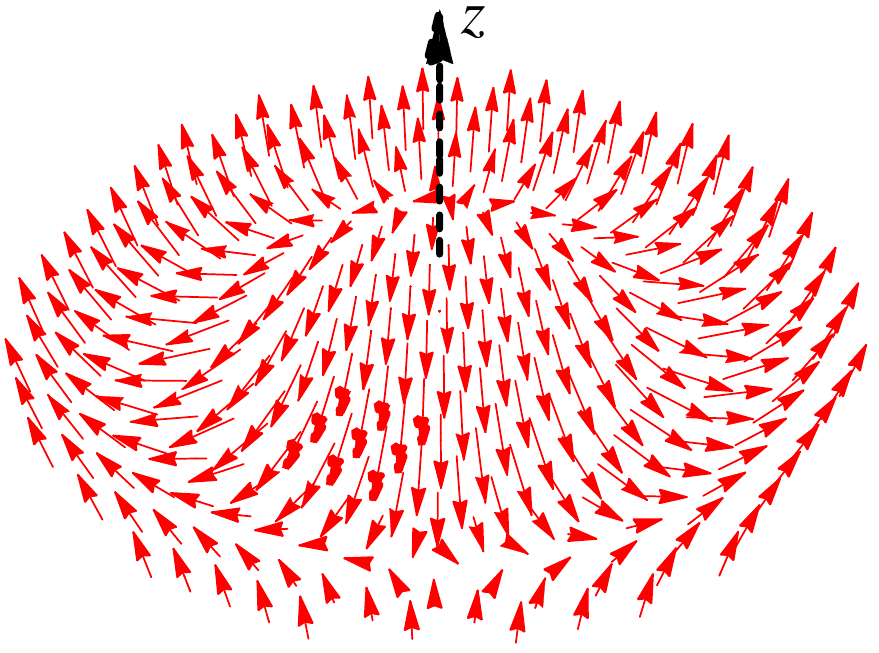}
	\caption{Structure of the \Neel vector $\bn_{\mathrm{sk}}(\mbr)$ of the static antiferromagnetic skyrmion with the profile plotted in figure \ref{fig:skyrmion_profile_09_plot}.}
	\label{fig:skyrmion_09}
\end{figure}

To calculate the constant $I$, we need to determine the profile of the antiferromagnetic skyrmion. The model in (\ref{eq:afm_free_energy}) allows for skyrmion solutions, as long as the external field is applied along the $\ez$-direction or is zero \cite{Bogdanov2002, Ivanov1995}. We assume the external magnetic field to be the dominant contribution towards $\bh_{\mathrm{eff}}$, so that the current-induced contribution towards the effective field does not destroy the skyrmion (that is, $H \gg C_1 \Ms j/ 2 \gamma$). The skyrmion profile $\bn = \bn_{\mathrm{sk}} + \delta \bn$ is determined from the steady state of (\ref{eq:final_AFM_equation_of_motion}) in the absence of dissipative and damping terms
\begin{equation}
	\bn \times \funWn
	+ 
	\frac{2}{\exch}
	(\bn \times \bh_{\mathrm{eff}}) (\bn \cdot \bh_{\mathrm{eff}}) 
	= 
	\frac{ 4 \Ms}{ \gamma \exch } \dnt (\bn \cdot \bh_{\mathrm{eff}}).
\label{eq:full_skyrmion_profile_equation}
\end{equation}
Here, the static component $\bn_{\mathrm{sk}}$ solves the equation with both $j, \dnt= 0$, whereas the corrections $\delta \bn$ originate from a non-zero dynamic term $\dnt$ and the current-induced effective field $ \Ca \Ms / 2 \gamma (\ez \times \bj)$. We are considering slow skyrmion dynamics, therefore the corrections can be assumed small, $| \delta \bn | \ll | \bn_{\mathrm{sk}} |$. 

The profile equation needs to be solved numerically \cite{Bogdanov2002}. For the purposes of the present work, we restrict the further analysis to the static component $\bn_{\mathrm{sk}}$ (the corrections $\delta \bn$ will not lead to a qualitative change in the skyrmion velocity estimate). It is convenient to rewrite (\ref{eq:full_skyrmion_profile_equation}) into spherical coordinates for the \Neel vector, $\bn_{\mathrm{sk}} = \left( \sin \theta \cos \psi, \sin \theta \sin \psi, \cos \theta \right)$, and cylindrical coordinates for the spatial variables, $\mbr = \rho (\cos \phi, \sin \phi)$. For the crystallographic class under consideration, a skyrmion solution exists when the \Neel order parameter has the same azimuthal direction as the cylindrical spatial vector (that is $\psi = \phi$, see figure \ref{fig:skyrmion_09}) \cite{Bogdanov2002}. The corresponding equation becomes
\begin{eqnarray}
	\parThRhoRhot
	+
	\frac{1}{\rhot}
	\parThRhot
	-
	\frac{ 	\sin \theta \cos \theta 		}{ 	\rhot^2 	}
	+
	\frac{ 	4 \dmiD 		}{ 	\pi \dmiD_0 	}
	\frac{ 	\sin^2 \theta 	}{ 	\rhot 	 		}
	-
	\left(
		1 
		-
		\frac{ 	 \h^2 	 }{ 	 \h_0^2 	 }
	\right)
	\sin \theta \cos \theta 
	=
	0.
\label{eq:skyrmion_profile_equation}
\end{eqnarray}
Here $\xo = \sqrt{\iExch  / \left| \anis \right|}$ is the characteristic lengthscale (domain wall width) of the system, \mbox{$\h_0 = \sqrt{\left| \exch \anis \right|}$} the spin-flop field, $\dmiD_0 = 4/\pi \sqrt{\iExch \left| \anis \right|}$ the threshold value of the inhomogeneous DM interactions constant that stabilizes modulated magnetic structures at zero magnetic field and $\rhot = \rho / \xo$ the rescaled radial coordinate. 

\begin{figure}[t!]
	\centering
		\includegraphics[scale=0.5]{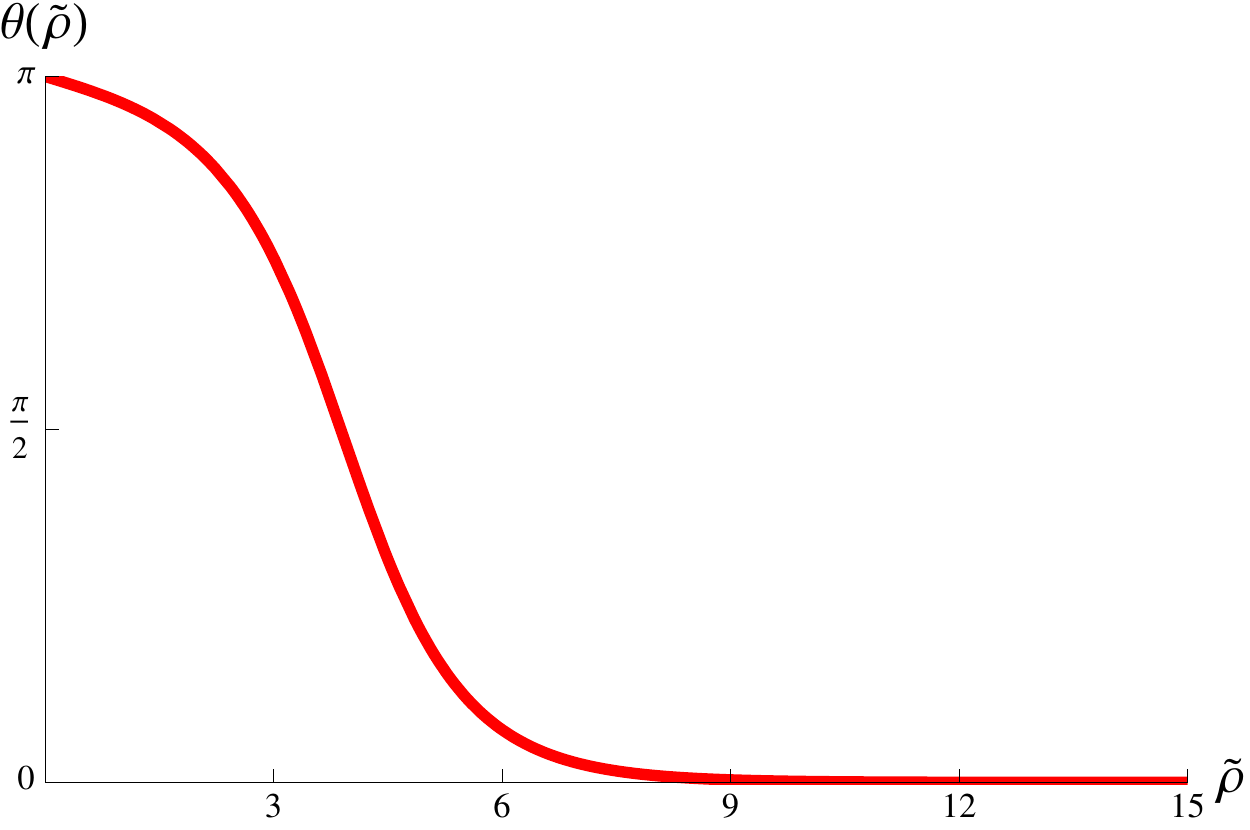}
	\caption{Antiferromagnetic skyrmion profile obtained by numerically solving (\ref{eq:skyrmion_profile_equation}) for $\h/\h_0 = 0.3$ and $\dmiD/\dmiD_0 = 0.9$.}
	\label{fig:skyrmion_profile_09_plot}
\end{figure}

Figure \ref{fig:skyrmion_profile_09_plot} shows the skyrmion profile for the particular choice of $\dmiD/\dmiD_0 = 0.9$ and $\h/\h_0 = 0.3$. The boundary conditions used are $\theta (\rhot = 0) = \pi$ and $\theta (\rhot \rightarrow \infty) = 0$, where the latter condition is implemented by a shooting method. With this, we are in a position to calculate numerically the integral $I$. The result is given in \ref{section:app_integrals}. 

\subsection{Skyrmion velocity}

Now, we are in a position to give an estimate for the magnitude of the longitudinal skyrmion velocity (\ref{eq:longitudinal_skyrmion_velocity}). We estimate the spin-orbit torque coefficient to be \mbox{$\Cb \simeq 3.4 \times 10^{-3} \, \mathrm{m^2 A^{-1} s^{-1}}$}  (see \ref{app:estimates} and \cite{Zelezny2014}). The Gilbert damping parameter is $\Gn \simeq 0.01$ \cite{Hals2011}. The characteristic length of the system is of the order of the skyrmion size, which is typically $\xo \simeq 10^{-8} \, \mathrm{m}$ \cite{Bogdanov2002, Hals2011, Tveten2013}. Typical experimentally used current densities in ferromagnets are of the order of $j \simeq 10^{11} \, \mathrm{A m^{-2}}$ \cite{Knoester2014}. 
With this, we estimate
\begin{eqnarray}
	v_{\mathrm{sk}, \Cb}
	\simeq
	255
	\,
	\mathrm{m s^{-1}}.
\end{eqnarray}
In contrast, \cite{Zhang2015b} predicts a skyrmion velocity of $\sim 1700 \, \mathrm{m / s}$. In that work, homogeneous torques of a similar form, but of different origin have been considered. The homogeneous torques there arise from a spin-polarized current injected vertically into the system, whereas the homogeneous torques in the present work are due to the spin-orbit coupling in the antiferromagnet. Using the values that they provide (see table \ref{tab:velocity_comparison}), we arrive at $v_{\mathrm{sk},\Cb} \simeq 500 \, \mathrm{m s^{-1}}$, which has the same order of magnitude as the result of \cite{Zhang2015b}.

In (\ref{eq:LLG_eqs_final}) the dissipative spin-transfer torque coefficient $\beta$ has dimensions of $\mathrm{ m^3 A^{-1} s^{-1}}$. Its dimensionless counterpart $\tilde{\beta}$ is obtained by $\tilde{\beta} = \beta n e$, where $n$ is the electron density and $e$ the electron charge. Typically, in ferromagnetic systems this value is taken to be of the order of the Gilbert damping, $\tilde{\beta} \simeq \Gn$ \cite{Zhang2004}. Typical metallic electron densities are of the order of $n \simeq 10^{29} \, \mathrm{ m^{-3} }$ and $e = 1.6 \times 10^{-19} \mathrm{As}$, so that for these parameters
\begin{eqnarray}
	v_{\mathrm{sk},\beta}
	&\simeq
	5
	\,
	\mathrm{m s^{-1}}.
\end{eqnarray}
This component of the velocity corresponds to the skyrmion velocity in \cite{Barker2015}. Our estimate agrees with the findings of that work for the same choice of parameters (see table \ref{tab:velocity_comparison}).

\begin{table}
\caption{\label{tab:velocity_comparison} Estimated skyrmion velocities using the parameters in the present work and in \cite{Zhang2015b, Barker2015}. In all cases the velocities $v_{\mathrm{sk},\Cb}$ and $v_{\mathrm{sk},\beta}$ are estimated according to the expressions in (\ref{eq:longitudinal_skyrmion_velocity}). The numerical results for $v$ are taken from \cite{Zhang2015b} and \cite{Barker2015}, respectively.} 
\begin{indented}
\footnotesize
\item[]
\begin{tabular}{@{}lllll}
\br		
Parameters			&Units 						&This work					&\cite{Zhang2015b}				&\cite{Barker2015}$^{\star \star}$					\\
\mr
$\Gn$				&$-$							&$0.01$						&$0.3$						&$0.01$									\\
$\Cb$				&$\mathrm{m^2 A^{-1}s^{-1} }$		&$0.0034 $		 			&$0.2^{\star}$					&$-$										\\
$\beta$				&$-$							&$0.01$						&$-$							&$0.1$									\\
$\xo$				&$\mathrm{m}$					&$10^{-8}$					&$10^{-8}$					&$-$										\\
$j$					&$\mathrm{A m^{-2}}$			&$10^{11} $					&$10^{11} $					&$3.2 \times 10^{12} \mbox{ }^{\star \star \star}$		\\
\mr
Estimated				&							&							&							&										\\
\mr
$v_{\mathrm{sk},\Cb}$	&$\mathrm{m s^{-1}}$				&$255$						&$500 $						&$-$										\\
$v_{\mathrm{sk},\beta}$	&$\mathrm{m s^{-1}}$				&$5$							&$-$							&$1000$									\\
\mr
Numerical results		&							&							&							&										\\
\mr
$v$					&$\mathrm{m s^{-1}}$				&$-$							&$\sim 1700$					&$\sim 2000$								\\ 
\br
\end{tabular}\\
$^{\star}$This value has been calculated from the expression that the authors provide in \cite{Zhang2015b} for their Slonczewski-like spin-transfer torque coefficient $\beta = \left| \hbar/ (\mu_0 e) \right| P / (2 d M_s)$, multiplied by the gyromagnetic ratio $\left| \gamma \right| = 2.211 \times 10^{5} \, \mathrm{m A^{-1}}$. Here, $\hbar$ is the reduced Planck constant, $\mu_0$ the vacuum permeability, $e$ the electron charge, $P=0.4$ is the polarization rate of the spin-polarized current, $d=0.4 \, \mathrm{nm}$ the film thickness and $M_s = 290 \, \mathrm{kA \,  m^{-1}}$ is the saturation magnetization; 
$^{\star\star}$For comparison, we focus only on one set of values for $\Gn, \beta$ and $v$ of the range provided in \cite{Barker2015};
$^{\star\star\star}$The authors use a value of $j = 200 \mathrm{m / s}$ for the drift velocity of the electrons. We calculate the corresponding current density by taking the electron density to be $n \simeq 10^{29} \mathrm{m}^{-3}$ and the electron charge $e = 1.6 \times 10^{-19} \mathrm{A s}$.

\end{indented}
\end{table}

\section{Conclusion and Outlook}

In summary, we extended the phenomenological theory of a uniaxial antiferromagnet with DM interactions to incorporate the current-induced spin-orbit torques together with the already studied spin-transfer torques. We used this theory to analyze the translational skyrmion motion in the presence of a time-independent external magnetic field and a DC electric current. We find that the magnetic field merely modifies the shape of the antiferromagnetic skyrmion and does not contribute towards the skyrmion motion. Further, our results show that the skyrmion moves in a straight line, along the direction of the applied electric current. This agrees with the numerical results of \cite{Zhang2015b, Barker2015}, which were obtained for skyrmions in the absence of a magnetic field. Depending on the choice of parameters, we find skyrmion velocities that are in the range of $1 - 1000 \, \mathrm{m s^{-1}}$, in agreement with the numerical results of \cite{Zhang2015b, Barker2015}.

Numerical simulations need to be performed in the presence of an external magnetic field to verify our analytical results. Another direction to proceed would be to allow for a variable skyrmion radius within the collective coordinate approach. The latter could uncover additional interesting physics. 


\section*{Acknowledgements}
H. V. and G. S. thank Yuta Yamane for helpful discussions. H. V. is a recipient of a DFG-fellowship through the Excellence Initiative by the Graduate School Materials Science in Mainz (GSC 266). R. D. is supported by the Stichting voor Fundamenteel Onderzoek der Materie (FOM), the European Research Council (ERC) and is part of the D-ITP consortium, a program of the Netherlands Organisation for Scientific Research (NWO) that is funded by the Dutch Ministry of Education, Culture and Science (OCW). \mbox{H. V.}, G. S., O. G. and J. S.  acknowledge support from the DFG Transregional Collaborative Research Center (SFB/TRR) 173 \lq\lq{}Spin+X – Spin in its collective environment\rq\rq{} and the Alexander von Humboldt Foundation.

\newpage

\appendix

\section{Collective coordinates}
\label{section:app_collective_coordinates}
We assume that the time dependence of the \Neel vector is given by \mbox{$\bn (\mbr, t) = \bn (\mbr - \bR(t))$}, where $\bR(t)$ is the skyrmion position (see section \ref{section:skyrmion_motion}). Within the collective coordinate approach we take $\xi_x = R_x$ and $\xi_y = R_y$ as the collective coordinates. Consequently, the partial derivatives appearing in (\ref{eq:collective_coordinates_ansatz}) need to be evaluated as
\begin{eqnarray}
	\frac{ \partial \bn (\mbr, t) }{ \partial \xi_i }
	=
	\frac{ \partial \bn (\mbr - \bR(t)) }{ \partial R_i }
	=
	- 
	\frac{ \partial \bn (\mbr)}{ \partial i },
\end{eqnarray}
for $i = x, y$.

\section{Integrals}
\label{section:app_integrals}
Here, we give the expression of the constant $I = I_2 / I_1$ that appears while transforming (\ref{eq:final_AFM_equation_of_motion}) into (\ref{eq:final_skyrmion_collective_coordinate_eom})
\begin{eqnarray}
	I_1		&=
				\intrhot
				\left[
					\left( 			\parThRhot		\right)^2
					\rhot
					+
					\frac{ \sin^2 \theta }{ \rhot }
				\right],
\nonumber \\
	I_2 		&= 
				\intrhot
				\left[
					\left( 			\parThRhot		\right)
					\rhot
					+
					\sin \theta \cos \theta
				\right].
\label{eq:app_list_of_integrals_to_calculate}
\end{eqnarray}
For the parameter choice $\dmiD/\dmiD_0 = 0.9$ and $\h/\h_0 = 0.3$, we evaluate it numerically to
\begin{eqnarray}
	I_1		&=		8.2,		\nonumber \\
	I_2 		&= 		-12.2,	
\label{app:integrals_09}
\end{eqnarray}
and, consequently, $I = -1.5$.

\section{Estimates}
\label{app:estimates}
Here, we estimate the spin-orbit torque coefficient $\Cb$. In \cite{Zelezny2014} the authors consider a torque of the form 
\begin{equation}
	\tau_{\bmA} 
	\propto
	\bmA \times \left[ \bmA \times \left( \ez \times \bj \right) \right],
\end{equation}
and give a numerical value of the field $\bBA \propto \bmA \times (\ez \times \bj)$ for interband processes. The value of that field for magnetization vectors pointing along the $\ez$-direction is found to be $\left| \bBA \right| \simeq 0.2 \, \mathrm{mT} $ per $0.1 \, \mathrm{A \,  cm^{-1}}$ so that 
\begin{equation}
	\left| \bBA \right| 
	\simeq
	2 \times 10^{-5} 
	\mathrm{ \frac{ T m }{ A } }
\end{equation}
From (\ref{eq:SOT_terms}) it follows that, in the corresponding torque $(\Cb / 2) \bM \times \left[ \bn \times (\ez \times \bj) \right]$, the coefficient has units 
$
	\left[ \Cb \right]
	=
	\left[
		\mathrm{ m^2  A^{-1} s^{-1} }
	\right].
$

Note the difference in dimensionality of $\left[ \bmA \right] = \left[ \mathrm{ A s^{-1}} \right]$ and $\left[ \bM, \bn \right] = \left[ 1 \right]$. The correct correspondence is
\begin{eqnarray}
	\frac{ \Cb }{ 2 }
	&\simeq
	\frac{\mu_B}{\hbar} \left| \bBA \right| 
	t 
	=
	1.7 \times 10^{-3} 
	\frac{ 
		\mathrm{ m^2 }
		}{ 
		\mathrm{A s }
		}.
\end{eqnarray} 
Here $t = 1$  $\mathrm{ nm }$ is a typical value for the thickness of a thin film. It has to be added to the calculation, because in \cite{Zelezny2014} the authors consider a strictly two-dimensional film. 

\section*{References}

\bibliography{AFMskyrmion}

\end{document}